\documentclass[conference]{IEEEtran}
\IEEEoverridecommandlockouts
\usepackage{cite}
\usepackage{amsmath,amssymb,amsfonts}
\usepackage{algorithmic}
\usepackage{graphicx}
\usepackage{textcomp}
\usepackage{xcolor}
\usepackage{hyperref}
\usepackage{caption}
\usepackage{subcaption}
\usepackage{lipsum,xcolor}
\usepackage{midfloat,capt-of}
\usepackage{multirow}
\usepackage{booktabs}
\usepackage{subfiles}
\usepackage[square,sort&compress,comma,numbers]{natbib}
\def\url#1{}
\hypersetup{colorlinks=true,
     linkcolor=black,
     filecolor=black,
     citecolor = black,      
     urlcolor=black,
     }

\def\BibTeX{{\rm B\kern-.05em{\sc i\kern-.025em b}\kern-.08em
    T\kern-.1667em\lower.7ex\hbox{E}\kern-.125emX}}
\begin{document}

\title{Comparison of Intelligent Approaches for Cycle Time Prediction in Injection Moulding of a Medical Device Product\\

\thanks{This research is supported by an IT Sligo Bursary and also by a research grant from Science Foundation Ireland (SFI) under Grant Number 16/RC/3872 and is co-funded under the European Regional Development Fund and by I-Form industry partners.}
}
\author{\IEEEauthorblockN{Mandana Kariminejad}
\IEEEauthorblockA{\textit{Centre for Precision Engineering,} \\
\textit{Materials and Manufacturing} \\
\textit{Institute of Technology Sligo}\\
Sligo, Ireland \\
mandana.kariminejad@mail.itsligo.ie}
\and
\IEEEauthorblockN{David Tormey}
\IEEEauthorblockA{\textit{Centre for Precision Engineering,} \\
\textit{Materials and Manufacturing} \\
\textit{Institute of Technology Sligo}\\
Sligo, Ireland  \\
Tormey.David@itsligo.ie}
\and
\IEEEauthorblockN{Saif Huq}
\IEEEauthorblockA{\textit{School of Computing and Digital Media} \\
\textit{London Metropolitan University}\\
London, United Kingdom \\
s.huq@londonmet.ac.uk}
 \and
 
\IEEEauthorblockN{Jim Morrison}
\IEEEauthorblockA{\textit{Department of Electronics and Mechanical Engineering} \\
\textit{Letterkenny Institute of technology}\\
Donegal, Ireland \\
jim.morrison@lyit.ie}
\and
\IEEEauthorblockN{Marion McAfee}
\IEEEauthorblockA{\textit{Centre for Precision Engineering, Materials and Manufacturing} \\
\textit{Institute of Technology Sligo}\\
Sligo, Ireland \\
McAfee.Marion@itsligo.ie}

}

\maketitle

\begin{abstract}
Injection moulding is an increasingly automated industrial process, particularly when used for the production of high-value precision components such as polymeric medical devices. In such applications, achieving stringent product quality demands whilst also ensuring a highly efficient process can be challenging. Cycle time is one of the most critical factors which directly affects the throughput rate of the process and hence is a key indicator of process efficiency.
In this work, we examine a production data set from a real industrial injection moulding process for manufacture of a high precision medical device. The relationship between the process input variables and the resulting cycle time is mapped with an artificial neural network (ANN) and an adaptive neuro-fuzzy system (ANFIS). The predictive performance of different training methods and neuron numbers in ANN and the impact of model type and the numbers of membership functions in ANFIS has been investigated. The strengths and limitations of the approaches are presented and the further research and development needed to ensure practical on-line use of these methods for dynamic process optimisation in the industrial process are discussed.

\end{abstract}


\begin{table*}[]
\centering
\caption{Comparison of the different training algorithm}
\label{I}
\resizebox{\textwidth}{!}{%
\begin{tabular}{@{}cccccc@{}}
\hline
\textbf{Training   Method}                & \textbf{Number of Epochs} & \textbf{Training MSE} & \textbf{Test MSE} & \textbf{Network MSE} & \textbf{R-value} \\ \hline
Bayesian Regularisation (trainbr)  & 1000                      & 0.002            & 0.065                & 0.012               & 0.997          \\
Levenberg-Marquart (trainlm)              & 14                        & 0.298             & 0.264                & 0.315               & 0.91          \\
Gradient Descent (traingd)                & 6                         & 104.47           & 99.78              & 104.05             & -0.031         \\
Gradient descent with momentum (traingdm) & 6                         & 88.76           & 92.79               & 88.811              & 0.25            \\
Scaled conjugated Gradient (trainscg)     & 11                        & 3.535             & 0.444                & 2.571               & 0.72          \\
One step   secant (trainoss)              & 32                        & 2.515            & 2.058                 & 2.058               & 0.744          \\ \hline
\end{tabular}%
}
\end{table*}

\section{Introduction}
Injection moulding has been developed to manufacture plastic components rapidly, in large volumes, and with high precision. A key objective in injection moulding process is to minimise the cycle time without affecting the part quality. A shorter cycle time results in higher throughput rate and thus efficiency improvement. However, the optimisation of the cycle time is a challenging task since a high number of process parameters should be monitored and controlled simultaneously. A method that can model the relationship between the input parameters and key responses pertaining to the cycle time and hence the process efficiency is highly desired.  \\
One of the most popular methods for predicting and mapping nonlinear relationships between input and response data is Artificial Neural Network (ANN) which has the ability of adaptive learning and real-time performance \cite{Pezeshki2019}. This method has been used in injection moulding for prediction of part quality factors such as warpage and shrinkage \cite{Ozcelik2006,Oliaei2016,Shen2007,Altan2010a}, quality optimization of particular components such as a bi-aspheric lens \cite{Bensingh2019}, prediction of optimal process variables like injection pressure, injection time, filling time \cite{Sadeghi2000} and temperature \cite{Khomenko2019} and prediction of the component weight \cite{Chen2008}. \\
ANN has been combined with Fuzzy logic to capture the benefits of both systems and overcome their drawbacks, so an Adaptive Neuro-Fuzzy Inference System (ANFIS) has emerged. It employs the linguistic and numerical language in Fuzzy logic and the nonlinear mapping in ANN. This method addresses the difficulty of defining fuzzy rules by experts, and it can build a fuzzy inference system based on a hybrid training method. Besides that,  ANFIS can overcome the uncertainties originating from different variables and experiment by defining fuzzy membership functions \cite{Pezeshki2019}. The other advantages are rapid response, rapid learning, and robustness \cite{Sahin2017}. The method has previously been applied to predict different aspects of the injection moulding process. For example, Hernandez et al. used this approach to predict the internal dimensions of an injection moulded component \cite{Hernandez2012} and it has also been used to find the optimal initial process setting \cite{Mok2002,He2000}.\\
The primary objective of this paper to develop an intelligent model that can predict the cycle time of an injection moulded medical device component based on the process variables. 
In this study, two different modelling approaches are compared. In section \ref{(II)}, an Artificial Neural Network (ANN) has been investigated using the feedforward back-propagation method with two hidden layers and six different training algorithms. The training algorithm yielding the best predictive performance is selected. Then, the effect of increasing the number of neurons on the performance has been investigated. In section \ref{(III)}, an Adaptive Neuro-Fuzzy Inference System (ANFIS) is studied. The influence of changing the number of membership functions and the Fuzzy controller type has been investigated. In the final section, these two approaches are compared and the challenges and limitations associated with the practical application of the methods in an industrial injection moulding process is discussed.\\

\section{Artificial Neural Network Approach}
\label{(II)}

The data set has been provided by an industrial partner from a commercial medical device production process and contains six hundred data points. The input variables are the mould temperature, injection pressure, and switch-over pressure from filling to packing phase. The output variable is the cycle time.

\subsection{Comparison of six training algorithms in two-hidden layers network}

\begin{figure}
\includegraphics[scale=0.45]{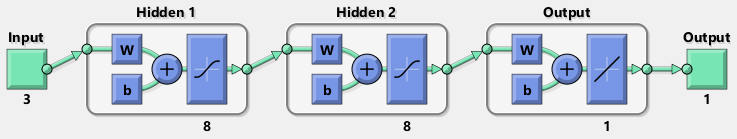}
\caption{Neural Network Structure}
\label{1}
\end{figure}

The feedforward back propagation ANN with two hidden layers and eight neurons has been used with six different training methods as: Bayesian regularization back-propagation (trainbr), Levenberg Marquardt (trainlm), Gradient Descent (traingd),  Gradient descent with momentum (traingdm), Scaled Conjugate Gradient (trainscg), and One Step Secant (trainoss). These training methods use different computation algorithms to optimise the performance of the network and minimise the mean square error. The data set has been divided into three parts; 70$\%$ for training, 15$\%$ for hyper parameter tuning and validation, and 15$\%$ for testing unseen data and evaluating the performance of the network.
The transfer function which has been used for hidden layers is Tangent Sigmoid defined in (\ref{(1)}): 

\begin{equation}
tansig(n)=\frac{2}{1+exp({-2}^n)}-1
\label{(1)}
 \end{equation}
This function is similar to tangent hyperbolic function, but it speeds up the computation in MATLAB software. A linear transfer function (purelin) has been implemented for the output layer. The schematic of the network with three inputs, two hidden layers, and eight neurons is illustrated in Fig. \ref{1}.

 The Mean Square Error (MSE) of training, test and the overall network performance has been compared for these six methods. The MSE error computes as  (\ref{(2)}), where ($y_{actual}$) is the actual cycle time and ($y_{predict}$)\ is the predicted one and $i$ is the number of iterations from 1 to $n$.

\begin{equation}
MSE=\frac{1}{n}\sum_{i=1}^{n}\left(y_i^{actual}-\ y_i^{predict}\right)^2
\label{(2)}
 \end{equation}

\begin{figure*}[pt]
    \centering
    \begin{subfigure}[b]{0.3\textwidth}
        \centering
        \includegraphics[width=\textwidth]{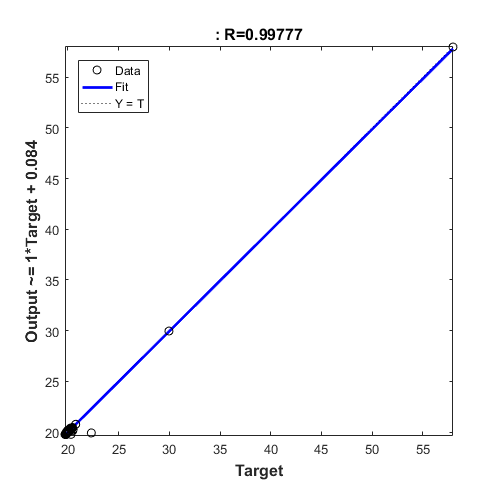}
        \caption{trainbr}
    \end{subfigure}
    \quad
    \begin{subfigure}[b]{0.3\textwidth}  
        \centering 
        \includegraphics[width=\textwidth]{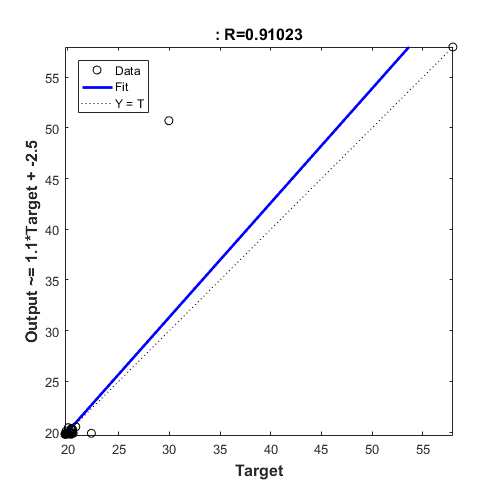}
        \caption{trainlm}
    \end{subfigure}
    \quad
    \begin{subfigure}[b]{0.3\textwidth}  
        \centering 
        \includegraphics[width=\textwidth]{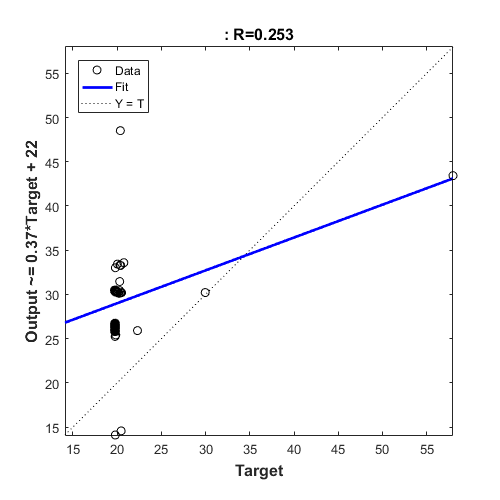}
        \caption{traingd}
    \end{subfigure}
    \vskip\baselineskip
    \begin{subfigure}[b]{0.3\textwidth}   
        \centering 
        \includegraphics[width=\textwidth]{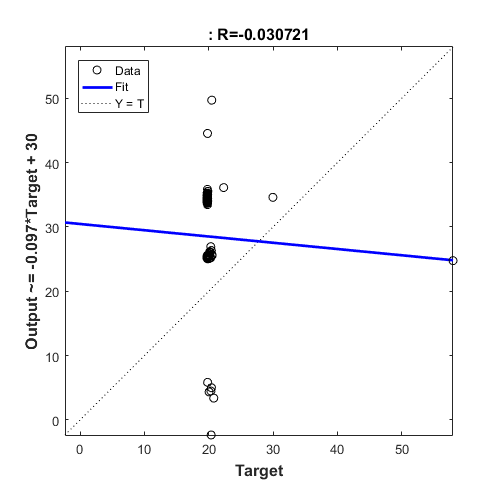}
        \caption{traingdm}
    \end{subfigure}
    \quad
    \begin{subfigure}[b]{0.3\textwidth}   
        \centering 
        \includegraphics[width=\textwidth]{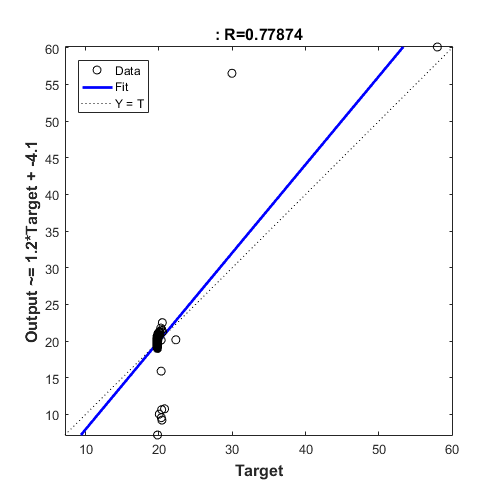}
        \caption{trainscg}
    \end{subfigure}
    \quad
    \begin{subfigure}[b]{0.3\textwidth}   
        \centering 
        \includegraphics[width=\textwidth]{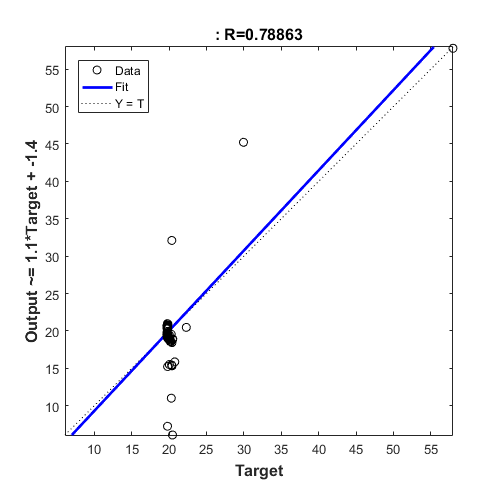}
        \caption{trainoss.}
    \end{subfigure}
    \caption{ANN regression plots for six training algorithms.}
    \label{2}
\end{figure*}

The quality of the regression model between the inputs and the cycle time has also been examined by the Pearson correlation coefficient R-value. Fig.  \ref{2}, shows the overall regression of different training algorithms. The summary of the performance of each training algorithm has been shown in Table \ref{I}.\\
As outlined in Table \ref{I}, the Bayesian regularisation back-propagation (trainbr) and Levenberg-Marquardt (trainlm) have the minimum MSE and the best performance. Besides that, the R-value is the maximum value for these two algorithms, indicating the data has been fitted well. Therefore these two training algorithms have been selected to assess the effect of neuron numbers.

\subsection{Effect of Number of Neurons}
\begin{figure}[h!]
    \centering
    \begin{subfigure}[b]{0.3\textwidth}
        \centering
        \includegraphics[width=\textwidth]{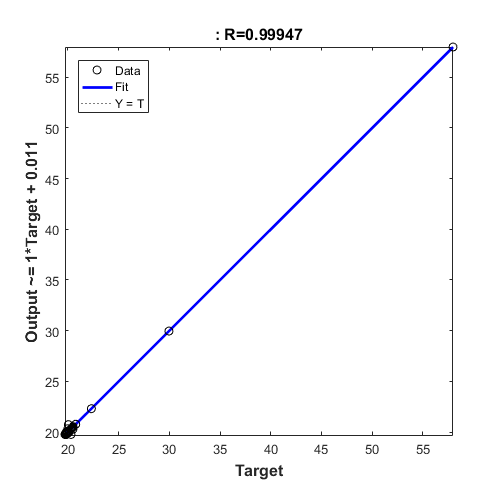}
        \caption{trainbr}
    \end{subfigure}
    \quad
    \begin{subfigure}[b]{0.3\textwidth}  
        \centering 
        \includegraphics[width=\textwidth]{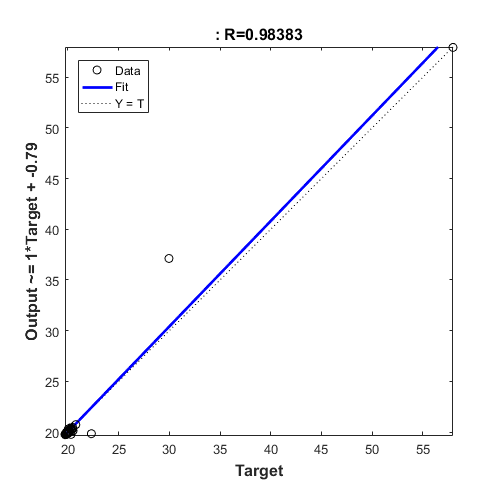}
        \caption{trainlm}
    \end{subfigure}
 \caption{ANN regression graphs for two hidden layers neural network with ten neurons}
    \label{3}
\end{figure}

To evaluate the effect of the number of neurons on the network, the number of neurons was increased from eight to ten  and  the Bayesian regularization back-propagation (trainbr) and Levenberg Marquardt (trainlm) have been compared together.\\
As a result of the increase in the number of neurons, the mean square error of the network for the Bayesian algorithm reduced from 0.012 to 0.003, and for the Levenberg-Marquardt from 0.315 to 0.098 with the R-value slightly increasing for both methods as shown in  Fig. \ref{3}.  Hence, the network performance is improved by the enhancement of neuron numbers. \\
In summary, in this section, the training of an ANN with six different training algorithms was investigated. The structure of the network is two hidden layers with eight neurons. The Bayesian regularization algorithm demonstrated the best result. The data is from a real industrial injection moulding process and thus contains uncertainties due to process disturbances and measurement noise. The Bayesian method explicitly considers this uncertainty in the model training. Levenberg-Marquardt training converges quickly,  after just fourteen iterations. However, the mean square error is not quite as low as that of the Bayesian method. The number of neurons was then increased from eight to ten for these two best-performing training algorithms. This improves the predictive performance, although it should be considered that an increase in the number of neurons also increases the computational time and cost. 

\section{Adaptive Neuro-Fuzzy Inference System (ANFIS) Approach}
\label{(III)}

\begin{figure}
\centering
\includegraphics[scale=0.6]{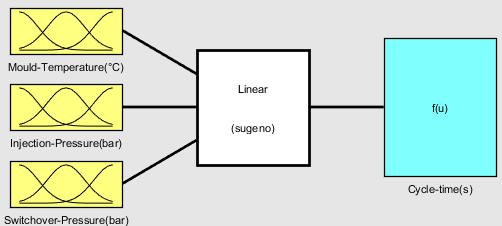}
\caption{Sugeno-based Fuzzy Inference System (FIS)}
\label{4}
\end{figure}

\begin{figure}
\centering
\includegraphics[scale=0.55]{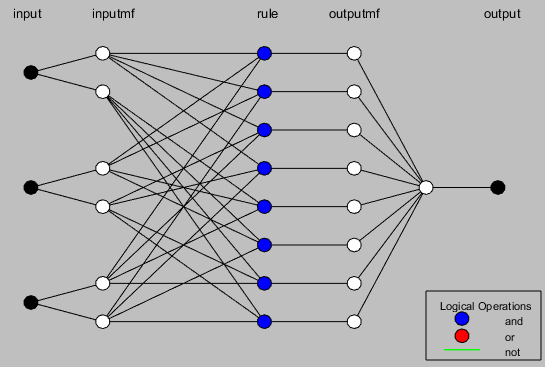}
\caption{ANFIS network structure with 2MFs.}
\label{5}
\end{figure}

The ANFIS method has been applied to fit the model between the inputs and the output and also evaluate the network performance by the unseen data. The data set has been divided into three parts. Four hundred data points have been used for training, a hundred for validation, and a hundred for testing.\\
A Gaussian Membership Function (MF) is applied for each input. The next step is specifying the number of membership functions (MFs); first, we examine the application of two MFs for each input and this is then increased to four to study the effect of the number of MFs. The Fuzzy Inference System (FIS) is Sugeno (See Fig. \ref{4}), and the impact of the change of Sugeno type from constant to linear (First order) is also investigated.\\
The structure of the ANFIS with two Membership Functions (MFs) for the inputs is illustrated in Fig. \ref{5}. The white nodes in this Figure are the adaptive nodes \cite{Jang1996}. The first layer makes the inputs fuzzy by defining the membership functions (MFS), and the hyperparameters of MFs should be tuned. For the Gaussian membership function with input value of $x$, the hyperparameters are the mean ($c$) and the standard deviation ($\sigma$), defined as (\ref{(3)}):

\begin{equation}
Gaussian(x,c,\sigma)=e^{\frac{-(x-c)^2}{2\sigma^2 }}
\label{(3)}
 \end{equation}

The second layer is based on the Fuzzy system rules, and the output of each node is the combination of the inputs of that node based on the defined rules in the fuzzy system. For example, for a system with two inputs ($x_1$, $x_2$), three MFs for each input in which $A_i$ are the MFs of $x_1$ and $B_i$ are the MFs of $x_2$ ($i$=1,2,3) and linear Sugeno fuzzy inference, the rules can be defined as \cite{Jang1996}:\\\\
Rule 1: If $x_1$ is $A_1$ And $x_2$ is $B_1$ Then $f_1$=$p_1$$x_1$+$q_1$$x_2$+$r_1$    \\                                                                     
Rule 2: If $x_1$ is $A_2$ And $x_2$ is $B_2$ Then $f_2$=$p_2$$x_1$+$q_2$$x_2$+$r_2$    \\                                                                          
Rule 3: If $x_1$ is $A_3$ And $x_2$ is $B_3$ Then $f_3$=$p_3$$x_1$+$q_3$$x_2$+$r_3$\\                                                                    

\begin{figure*}[pt]
    \centering
    \begin{subfigure}[b]{0.4\textwidth}
        \centering
        \includegraphics[width=\textwidth]{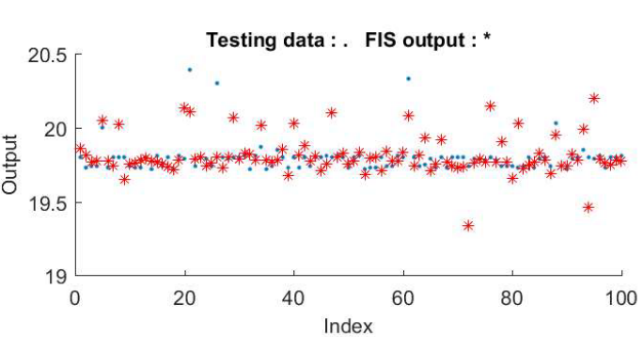}
        \caption{2MFs- Constant Sugeno}
    \end{subfigure}
    \quad
    \begin{subfigure}[b]{0.4\textwidth}  
        \centering 
        \includegraphics[width=\textwidth]{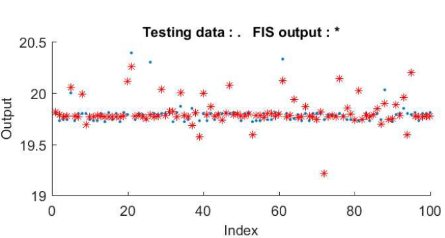}
        \caption{4MFs- Constant Sugeno}
    \end{subfigure}
    \vskip\baselineskip
    \begin{subfigure}[b]{0.4\textwidth}   
        \centering 
        \includegraphics[width=\textwidth]{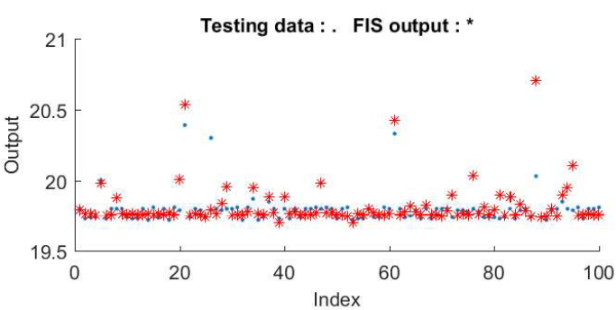}
        \caption{2MFs- Linear Sugeno}
    \end{subfigure}
    \quad
    \begin{subfigure}[b]{0.4\textwidth}   
        \centering 
        \includegraphics[width=\textwidth]{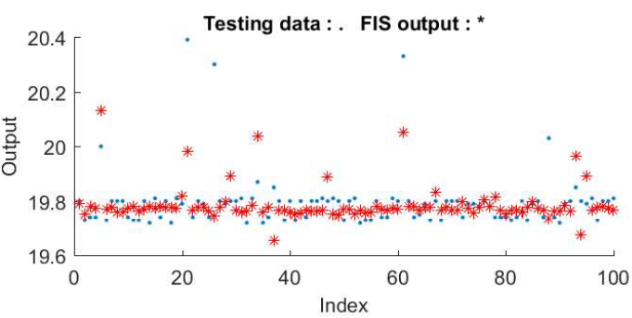}
        \caption{4MFs- Linear Sugeno}
    \end{subfigure}
    \caption{ANFIS result for test data with different MFs and Sugeno types.}
    \label{6}
\end{figure*}

The number of rules is the product of the number of MFs and the number of input variables. In a system with two inputs and three MFs, there would be six rules (here just three rules have been shown for illustration purposes). Any T-norm operator, which is a demonstrator of ''AND'', can be used to combine the MFs. In this study 'Product' function has been used and for example, the first rule ''If $x_1$ is $A_1$ And $x_2$ is $B_1$'' can be defined as $A_1$ $\times$ $B_1$.\\
The fourth layer is adaptive again, and the Sugeno parameters, defined in the rules, will be tuned there. The last layer obtains the desired output by summation of the inputs.\\
The method for generating the FIS network and classifying the data in this study is grid partitioning, which creates a Sugeno-type FIS (explained above) as initial conditions for ANFIS training. The next step is defining the neural network algorithm and the hybrid method selected to find the optimal error and speed up the process, a combination of Back Propagation and Gradient Descent methods \cite{Kumar2013}.\\
The summary of the training and testing MSE based on the number of MFs and the Sugeno type is presented in Table \ref{II}. Fig. \ref{6} shows the network performance for the test data.  The X-axis shows the number of the data point and the output on the Y-axis is the value of the cycle time (s). The red asterisks and the blue dots are the predicted value from the network and the actual cycle time value respectively.

\begin{table}[]
\centering
\caption{Summary of ANFIS}
\label{II}
\resizebox{\columnwidth}{!}{%
\begin{tabular}{cccc} \hline
\textbf{Number of   MFs} & \textbf{Sugeno type} & \textbf{Training MSE} & \textbf{Testing MSE} \\\hline
\multirow{2}{*}{2}       & Constant             & 0.111               & 0.132              \\
                         & Linear               & 0.057                & 0.11              \\\hline
\multirow{2}{*}{4}       & Constant             & 0.087              & 0.127              \\
                         & Linear               & 0.048              & 0.052      \\\hline      
\end{tabular}%
}
\end{table}

Comparison of the results in Table \ref{II} shows that by increasing the number of MFs, the training MSE and test MSE have been reduced slightly with both constant and linear Sugeno. Through  keeping the number of MFs the same and changing the Sugeno type from constant to linear, the errors have also shown to decrease. For instance, for the ANFIS network with 4MFs, the testing and training MSE is almost halved. Hence, increasing the number of MFs and changing the Sugeno type from constant to linear will both improve the network performance, while the  effect of the Sugeno type on the network error is more considerable.\\ It should be also considered that the increase of MFs and Sugeno type to linear means an increased the number of hyperparameters that should be tuned and a rise in the computation time and cost.

\section{Conclusion}

In this paper, the prediction of cycle time in the injection moulding process and its relationship with three input variables (mould temperature, injection pressure, and switch-over pressure) have been studied with ANN and ANFIS. For the ANN method comparing six different training algorithms, the Bayesian regularization had the best performance but the response is slow to train at 1000 iterations ( See Table \ref{I}).  Increasing the number of neurons reduced the prediction error. A minimum error of 0.003 has been achieved with ANN by Bayesian regularisation training with a two-hidden layer network with ten neurons. The ANFIS results showed that the network performance could be improved by increasing the number of MFs and changing the Sugeno model order. The minimum error obtained with ANFIS is 0.052 with first-order Sugeno and four membership functions.\\
By comparison of these two methods, the ANN with the Bayesian regularisation algorithm had the best performance. However, it should be mentioned both methods had error results below 0.2s which is acceptable variation for the cycle time of this injection moulded part. It should also be noted that the ANN cannot guarantee convergence and it is possible that even after a long learning process, the model diverges as occurred in this study with gradient descent algorithms (see Fig. \ref{2} $\&$ Table \ref{I} ). Divergence problems were not observed with the ANFIS method. It should also be considered that increasing the number of parameters in both approaches (by increasing the neuron numbers, MFs and switch to a more complex Sugeno type) will lead to increased computation complexity and increased time and cost.\\
This work highlights the potential of intelligent approaches for accurate prediction of cycle times in an industrial injection moulding process. Such models can be exploited within a process control scheme to maximise the process efficiency without compromising product quality, which is critical for industrial competitiveness in high-value, high-throughput manufacturing.
Further research will involve investigation of the influence of other process settings on the cycle time, a study of the impact of layer numbers in ANN and the type of MFs and training method in ANFIS. Finally the comparison of these methods with other machine learning approaches within a wider process control strategy will be further investigated.

\footnotesize{
\bibliographystyle{IEEEtran}
\bibliography{ANFISP.bib}
}

\end{document}